# Empirical modeling of Radiative *versus* Magnetic flux for the Sun-as-a-Star

**Dora Preminger[1] · Dibyendu Nandy[2] · Gary Chapman[1] · Petrus C.H. Martens[3]**

**Abstract**

We study the relationship between full-disk solar radiative flux at different wavelengths and average solar photospheric magnetic-flux density, using daily measurements from the Kitt Peak magnetograph and other instruments extending over one or more solar cycles. We use two different statistical methods to determine the underlying nature of these flux–flux relationships. First, we use statistical correlation and regression analysis and show that the relationships are not monotonic for total solar irradiance and for continuum radiation from the photosphere, but are approximately linear for chromospheric and coronal radiation. Second, we use signal theory to examine the flux–flux relationships for a temporal component. We find that a well-defined temporal component exists and accounts for some of the variance in the data. This temporal component arises because active regions with high magnetic field strength evolve, breaking up into small-scale magnetic elements with low field strength, and radiative and magnetic fluxes are sensitive to different active-region components. We generate empirical models that relate radiative flux to magnetic flux, allowing us to predict spectral-irradiance variations from observations of disk-averaged magnetic-flux density. In most cases, the model reconstructions can account for 85 – 90 % of the variability of the radiative flux from the chromosphere and corona. Our results are important for understanding the relationship between magnetic and radiative measures of solar and stellar variability.

# 1. Introduction

Solar magnetic fields are manifested as distinct features on solar images because their presence causes changes in the radiative properties of the solar atmosphere. The features with the strongest magnetic fields are known as active regions. At low, photospheric levels, active regions may include both dark sunspots with very strong magnetic fields and bright faculae with weaker fields, and therefore the average photospheric intensity of an active region depends on its average magnetic flux in a complex way (*e.g.* Ortiz and Rast, 2005). However, active-region intensity at chromospheric and coronal heights generally increases monotonically with increasing magnetic flux, presumably because the presence of the magnetic fields causes non-radiative energy to be channeled into the Sun's outer atmospheric layers (*e.g.* Klimchuk, 2006).

---

[1] San Fernando Observatory, California State University, Northridge, CA 91330, USA
e-mail: dora.preminger@csun.edu

[2] Department of Physical Sciences, Indian Institute of Science Education and Research, Kolkata, Mohanpur 741252, West Bengal, India

[3] Harvard-Smithsonian Center for Astrophysics, Cambridge, MA 02138, USA; Physics Department, Montana State University, Bozeman, MT 59715, USA

While emission from the solar chromosphere and corona changes dramatically over the course of the 11-year solar cycle, the average magnetic-flux density of an active region does not appear to change (Schrijver, 1987); therefore it would seem that the solar radiative variability is driven by the temporal variation in the *amount* of magnetic flux emerging on the solar surface (Schrijver and Harvey, 1989). It has been argued that if the only difference between solar maximum and solar minimum is the number of active regions, and all active regions are essentially identical, then we should expect the emission from the entire outer solar atmosphere to depend linearly on the total solar magnetic flux, since two active regions should be twice as bright as one, *etc*. Following this line of reasoning, we would also expect total coronal emission to be linearly related to total chromospheric emission.

The relationships between the radiative and magnetic-flux densities for the sun (referred to as the flux–flux relationships) have been studied in order to test this prediction. Several studies have shown that the flux–flux relations may not be linear at all. Pevtsov *et al*. (2003) have compared full-disk solar X-ray flux to full-disk unsigned magnetic flux. Using almost ten years of soft X-ray data from Yohkoh and full-disk magnetograms from Kitt Peak, they found the relationship was a power law with an index between 1.6 and 2.0. However, they found a linear relationship for discrete solar features, as for other active stars. We have not been able to find solar studies comparing full-disk chromospheric flux to full-disk photospheric magnetic flux. Livingston *et al.* (2007) discuss the spectrum variability of the Sun-as-a-star and its correlation with the sunspot number. Schrijver and Harvey (1989) and Harvey and White (1999) found that average Ca II K-line flux density is proportional to average magnetic-flux density to the power ≈ 0.6, when averaging over non-sunspot chromospheric regions. Ortiz and Rast (2005) obtained a similar result when comparing Ca II K-line intensity with magnetic-flux density for image sections near disk center. Also, Schrijver *et al*. (1983) showed that the average solar coronal X-ray flux depends on the average chromospheric flux to the power ≈1.6.

Non-linear flux–flux relationships are obtained both for the Sun and other Sun-like stars, a fact that is not well understood (Haisch and Schmitt, 1996). It appears to be inconsistent with the idea that an increase in activity level is caused by an increased number of "standard" active regions. It may indicate that the structure of active regions actually *is* a function of activity level, or it may be a result of the particular way in which surface magnetic flux emerges and evolves. After appearing on the surface, active regions decay, partly disappearing *in situ*, and partly dispersing into the diffuse, low-contrast magnetic network (Schrijver and Harvey, 1989). Small-scale magnetic regions such as ephemeral regions also contribute to the magnetic flux in the network. Schrijver (2001) developed a model with a flux-dependent rate of field dispersal in order to explain the non-linear relation between the chromospheric emission and magnetic-flux density. But recently, Preminger and Walton (2007) have shown that scatter plots of parameters that describe different parts of an active region are often non-linear simply because of active-region evolution.

Given the large amount of solar data currently available, it seems appropriate to re-examine the flux–flux relationships for the Sun, particularly the globally-averaged relations, which are expected to be applicable to other magnetically active, cool stars. In the study of cool stars, radiative flux is often used as a proxy for magnetic activity, thus it would be helpful to understand in more detail how radiative flux is related to magnetic flux. Such a study revisiting the flux–flux relationship might also throw light on the nature of magnetic heating and radiation from stellar atmospheres, by providing empirical relationships that set quantitative constraints on theoretical models of magnetic heating (*e.g.* Fisher *et al*., 1998). In this paper, we study flux–flux relations for the Sun-as-a-star, using synoptic disk-integrated observations that have been taken almost daily for one or more solar cycles. We compare long sequences of daily data for magnetic-flux density and various measures of solar radiative flux, at several different wavelengths. The use of large data sets allows us to use statistical methods to analyze the relationships.

## 2. Data

For this study we need a synoptic full-disk measurement of photospheric magnetic-flux density, $|B|$, which has reasonably good resolution and is available for a long period of time. We choose $|B_{KP}|$, the average unsigned magnetic-flux density from Kitt Peak, in Gauss (G). This is the absolute value of the line-of-sight magnetic-field strength observed with 1 arcsec pixel size, averaged over the full disk. The Kitt Peak data are based on daily magnetograms available 1977 – 2003, covering most of Solar Cycles 21–23 and are shown in the top panel of Figure 1. These data are widely-used and have been well-studied. Note that $|B_{KP}|$ is a composite data set: the data prior to 1992, obtained with the 512-channel magnetograph, have a noise level of about 8 G, while data obtained after 1992, with the spectromagnetograph, have a lower noise level of about 5 G (Wenzler *et al*., 2005, 2006). The data suffer from uncertainties inherent in any long-term composite. Following Wenzler *et al*. (2006), we have corrected the data prior to 1990 by multiplying by a factor of 1.242. This correction was shown by Wenzler to improve the consistency of the composite, but uncertainties remain. Uncertainties are also introduced by observational artifacts in the magnetograms; the quality of the data is not consistent from day to day, and some days are missing. In addition, since the magnetograms measure line-of-sight magnetic-field strength, $|B_{KP}|$ underestimates the true magnetic flux in regions close to the limb. Because this is a statistical study, we consider these uncertainties to constitute a source of noise. Where data gaps are present (≈30% of days), we interpolate the data using a shape-preserving piecewise cubic interpolation.

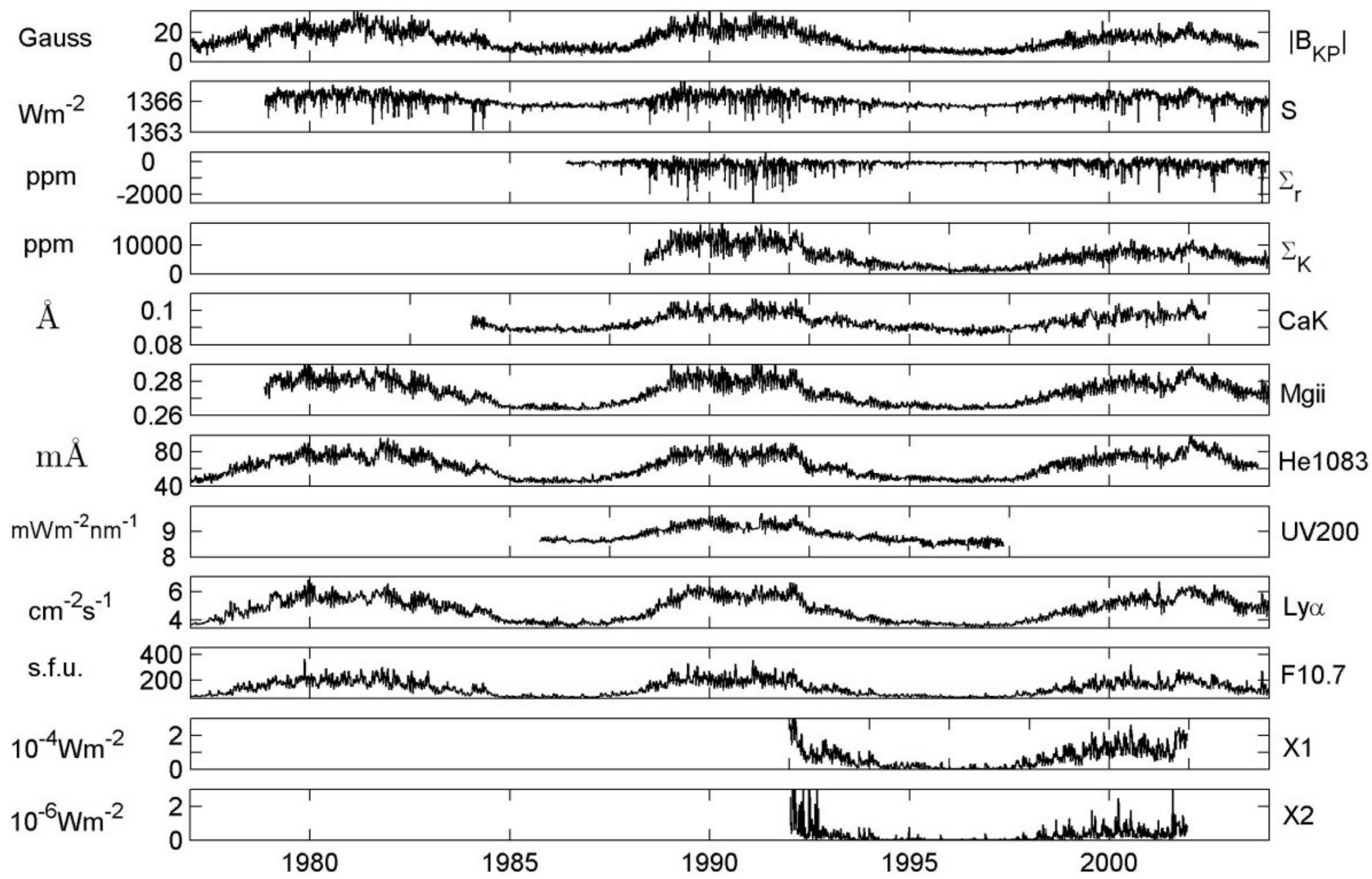

**Figure 1:** Top panel: $|B_{KP}|$, the average unsigned magnetic-flux density from Kitt Peak, in Gauss. Lower panels: The radiative-flux measurements [$F(\lambda)$] studied in this work. Details of the $F(\lambda)$ are given in Table 1.

It is important to clarify exactly what $|B_{KP}|$ measures. Formally, it is the disk-averaged magnetic-flux density measured in units of Gauss; however, since the disk-area of the Sun stays the same, this quantity $|B_{KP}|$ may in fact be interpreted as a proxy for the total magnetic flux on the visible solar disk (Harvey and White, 1999). $|B_{KP}|$ is not equal to true $|B|$, because it underestimates the flux contributions of certain active-region components. The magnetograms significantly underestimate sunspot magnetic fields, by up to a factor of two (Wenzler *et al*., 2004), so sunspot flux is under-represented in $|B_{KP}|$. $|B_{KP}|$ is also insensitive to small-scale magnetic features, because of the resolution limitation inherent in the magnetograms due to seeing and small-scale flux cancellation. Krivova *et al*. (2002) and Krivova and Solanki (2004) estimate that apparent cancellation of small-scale flux causes about half the flux in small-scale magnetic elements to go undetected in Kitt Peak synoptic charts. These limitations need to be taken into account when interpreting the results of our study.

**Table 1.** Disk-integrated observations and proxies of solar radiative flux at different wavelengths [$F(\lambda)$].

| Symbol | Observable | Units | λ(nm) | Solar Origin | Data Source(s) | Time period |
|---|---|---|---|---|---|---|
| S | Total Solar Irradiance at 1 AU | W $m^{-2}$ | all | all layers of solar atmosphere | Satellite Composite from PMOD/WRC, (Fröhlich, 2000; 2006) http://www.pmodwrc.ch/ | 1978 – present 93% coverage |
| $\Sigma_r$ | Change in red continuum intensity due to features | ppm of quiet Sun | 672.3 | Photosphere | Broadband red images from San Fernando Observatory http://www.csun.edu/sfo/ | 1986 – present 61% converge |
| $\Sigma_K$ | Change in Ca II K-line intensity due to features | ppm of quiet Sun | 393.0 | Upper photosphere and lower chromosphere | Ca II K-line images (1 nm bandpass) from San Fernando Observatory, http://www.csun.edu/sfo/ | 1988 – present 61% coverage |
| CaK | Equivalent width | Å | 393.4 | chromosphere | Emission index from NSO / Sac Peak http://nsosp.nso.edu/data/cak_mon.html | 1984 – present 42% coverage |
| Mg II | Core / Wing Ratio | - | 280 | chromosphere | Satellite Composite by Viereck (2001), http://www.sec.noaa.gov/ftpmenu/sbuv.html | 1978 – present 100% coverage |
| He1083 | Equivalent width | mÅ | 1083 | chromosphere | Spectra from NSO / Kitt Peak http://nsokp.nso.edu/ | 1975 – present 100% coverage |
| UV200 | Ultraviolet Irradiance at 1 AU | mW $m^{-2}$ $nm^{-1}$ | 200 – 205 | chromosphere | NOAA-9 SBUV/2 Calibrated by DeLand (2004), http://ozone.sesda.com/solar/ | 1985 – 1997 92% coverage |
| Ly α | Lyman-α irradiance at 1 AU | photons $cm^{-2}$ $s^{-1}$ | 121.6 | chromosphere and transition region | Satellite Composite by Woods (2000), http://lasp.colorado.edu/solstice/data.html | 1977 – present 100% coverage |
| F10.7 | Radio Flux at 1 AU | s.f.u. $\equiv 10^{-22}$ W $m^{-2}$ $Hz^{-1}$ | 10.7 cm | upper chromosphere /lower corona | Ottawa / Penticton radio telescopes, http://www.ngdc.noaa.gov/stp/SOLAR/ftpsolarradio.html | 1947 – present 100% coverage |
| X1 | Daily average X-ray flux at 1 AU | W$m^{-2}$ | 0.8 – 2 | corona | X-ray Images from *Yohkoh* Soft X-ray telescope (SXT), http://solar.physics.montana.edu/ylegacy/irradiance | 1992 – 2001 90% coverage |
| X2 | Daily average X-ray flux at 1 AU | W$m^{-2}$ | 0.2 – 0.8 | corona | X-ray Images from SXT, http://solar.physics.montana.edu/ylegacy/irradiance | 1992 – 2001 90% coverage |

In this work we compare $|B_{KP}|$ to either observations or proxies of solar radiative flux at different wavelengths [$F(\lambda)$]. We choose disk-integrated data that have been observed daily for at least one solar cycle. Data gaps in $F(\lambda)$ are also filled in using a piecewise cubic interpolation. Table 1 describes these observations, their wavelengths, and the layer(s) of the solar atmosphere from which the radiative flux originates, along with the data sources, and the time-intervals for which the observations are available.

Figure 1 shows $|B_{\mathrm{KP}}|$ and the 11 temporal series $[F(\lambda)]$ being considered in this study. All but one of the $F(\lambda)$ vary in phase with $|B_{\mathrm{KP}}|$ on solar-cycle timescales. The exception is $\Sigma_{\mathrm{r}}$, the variability of red-continuum radiation from magnetic features in the photosphere. As discussed in Preminger *et al*. (2002), this quantity shows strong variability on solar-rotation timescales due to sunspots, but it is slightly anti-correlated with $|B_{\mathrm{KP}}|$ on solar-cycle timescales.

# 3. Exploring the Relationships between Disk-Averaged Radiative Flux and Magnetic-Flux Density

## 3.1 Statistical Correlation and Regression Analysis

Given co-temporal observations of $F(\lambda)$ and $|B_{\mathrm{KP}}|$, a straightforward way to investigate their relationship is to view and analyze a scatter plot of the data. Figure 2 shows sample scatter plots for select $F(\lambda)$ *vs.* $|B_{\mathrm{KP}}|$, showing all available data points. On the plots, each data point represents a pair of observations at a given instant in time, a regression analysis of which may reveal the empirical relationship between the two observed variables.

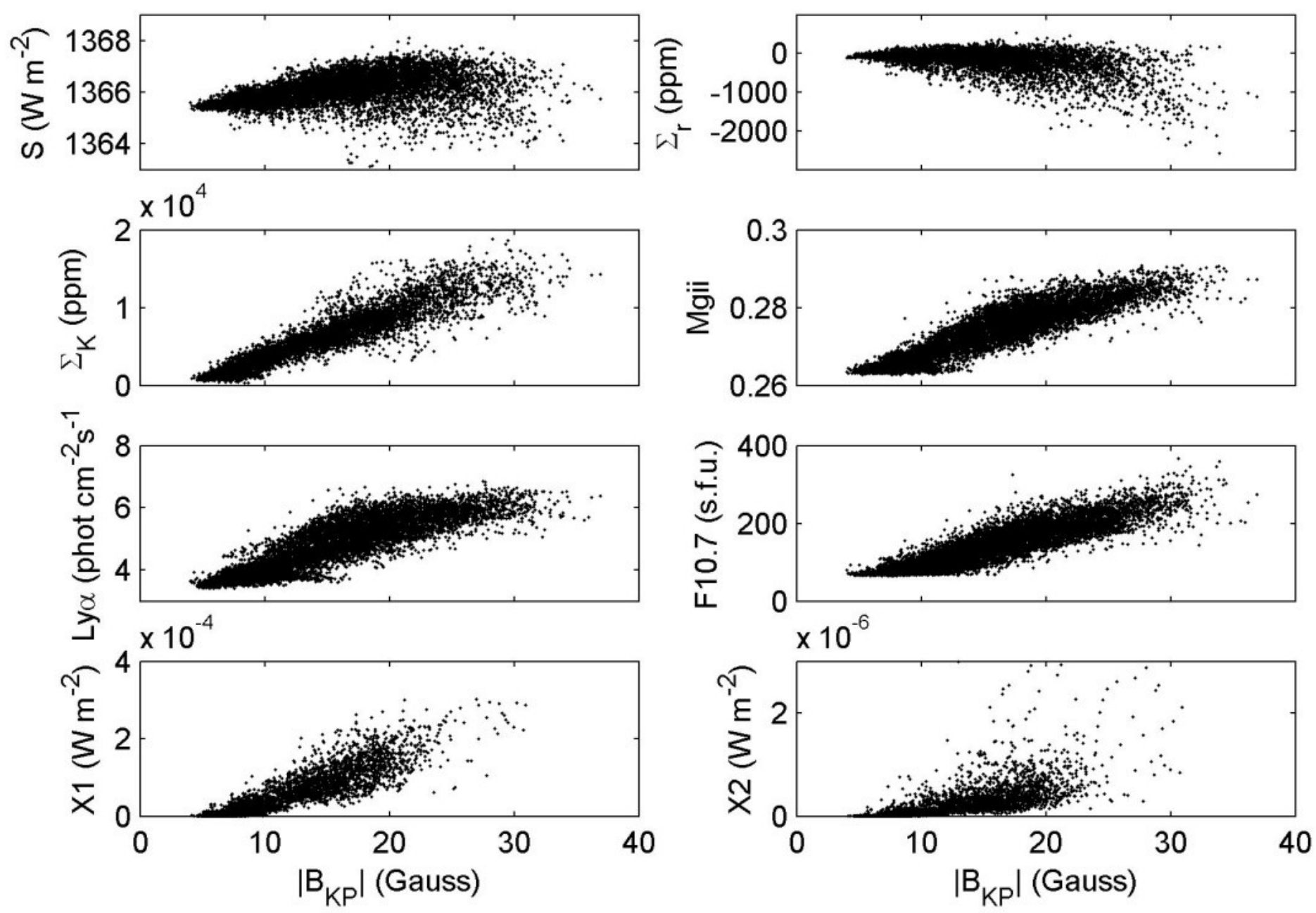

**Figure 2:** Scatter plots for select radiative flux measures: $F(\lambda)$, *vs.* $|B_{\mathrm{KP}}|$.

Table 2 quantifies the correlations that exist between $F(\lambda)$ and $|B_{\mathrm{KP}}|$when all available co-temporal data points are considered. The confidence levels for all the correlations are over 99%. We have computed both the Pearson's correlation coefficient and the Spearman's rank correlation coefficient. The Pearson test measures the strength of a linear correlation between paired measurements, while the Spearman test evaluates only whether there is a monotonic relationship between the variables. If the Spearman coefficient is significantly higher than the Pearson coefficient, this may indicate that the relationship between the variables is monotonic but not linear; or, alternatively, it may indicate that the data set(s) contains a large number of outliers, since these affect the Pearson coefficient more than the Spearman coefficient (Bernstein, 1999).

**Table 2** Quantifying the relationships between $F(\lambda)$ and $|B_{KP}|$. The scatter plots are almost linear except in the case $F(\lambda) = S$ or $\Sigma_r$. The FIR model for $F(\lambda)$ accounts for more of the variance than does the linear regression model because it allows for a temporal component to the relationship. The correlation coefficient for the FIR model is the Pearson correlation coefficient between the actual data and the modeled data generated through the FIR function.

| $F(\lambda)$ | $F(\lambda)$ vs. $|B_{KP}|$ Correlation and regression analysis | | | | FIR model fit (Eq. (3)) | | |
|---|---|---|---|---|---|---|---|
| | # data points | Pearson coeff | Spearman coeff | Best fit | Pearson coeff | $F_0(\lambda)$ | $a$ |
| S | 9075 | 0.43 | 0.54 | - | 0.45 | 1365.39 | 1.22 |
| $\Sigma_r$ | 6328 | - 0.49 | - 0.36 | - | 0.64 | 47 | 1.03 |
| $\Sigma_K$ | 5610 | 0.92 | 0.94 | $F = 546\ |B_{KP}|$ -1860 | 0.95 | -380 | 1.01 |
| CaK | 6721 | 0.89 | 0.89 | $F = 6.3 \times 10^{-4}\ |B_{KP}| + 0.0846$ | 0.91 | 0.0864 | 0.95 |
| Mg II | 9085 | 0.92 | 0.92 | $F = 9.95 \times 10^{-4}\ |B_{KP}| + 0.2581$ | 0.94 | 0.2610 | 1.15 |
| He1083 | 9754 | 0.86 | 0.88 | $F = 1.76\ |B_{KP}| + 36.5$ | 0.88 | 41.2 | 1.21 |
| UV200 | 4236 | 0.92 | 0.91 | $F = 0.0381\ |B_{KP}| + 8.34$ | 0.94 | 8.46 | 1.59 |
| Ly$\alpha$ | 9754 | 0.89 | 0.90 | $F = 0.1191\ |B_{KP}| + 2.94$ | 0.92 | 3.24 | 1.29 |
| F10.7 | 9754 | 0.91 | 0.91 | $F = 8.09\ |B_{KP}| + 15.1$ | 0.92 | 41.2 | 1.15 |
| X1 | 3631 | 0.88 | 0.91 | $F = 1.1 \times 10^{-5}\ |B_{KP}| - 6.7 \times 10^{-5}$ | 0.92 | $3.8 \times 10^{-5}$ | 1.05 |
| X2 | 3631 | 0.53 | 0.87 | $F = 5.1 \times 10^{-8}\ |B_{KP}| - 3.7 \times 10^{-7}$ | - | - | - |

As Table 2 shows, both the Pearson and the Spearman coefficients are low in the case of S and $\Sigma_r$, indicating that total solar irradiance and changes in the red continuum flux are not monotonic functions of $|B_{KP}|$. For all of the chromospheric and coronal fluxes, except for X2, the Pearson and Spearman coefficients are significant, and almost equally high, so these can be modeled as linear functions of $|B_{KP}|$.

For the shortest X-ray wavelengths [X2] the Spearman coefficient is significantly higher than the Pearson coefficient, so we explored the possibility of a quadratic or a power-law relationship between the variables. Goodness-of-fit tests show that these functions do not improve the fit quality (see Table 3 and discussion below). However, the scatter plot does reveal a large number of outliers (see Figure 2), and this can explain why the Pearson coefficient is lower than the Spearman. For these reasons, we conclude that the relationship between these variables is best described as linear.

Note that most of the plots in Figure 2 include data from multiple solar cycles. The correlation analysis described above implicitly assumes that the relationship between $F(\lambda)$ and $|B_{KP}|$ is the same for all cycles. In Table 3, we compare the regression coefficients [$R^2$] for a linear fit to the data for each solar cycle as well as for all data points. The regression coefficient indicates what percentage of the variance is accounted for by the model. It is clear that in most cases the linear fits are best when cycle 22 is considered alone. When data for all cycles are included, the differences that exist between cycles result in increased variance in the scatter plots.

The linear relations shown in Table 2 can be used to reconstruct solar spectral irradiance for all times for which $|B_{KP}|$ is known. We can compare the reconstructions with existing observations and study the fit residuals, to better evaluate the validity of the linear model. While the regression coefficients indicate that the linear correlation between the variables is highly significant, analysis of the $\chi^2$ goodness-of-fit statistic for the fit residuals shows that these residuals do not have a normal distribution. A quantile–quantile plot of the residuals is somewhat S-shaped, the residuals being anomalously high (low) at high (low) values of $F(\lambda)$. This means that there is a pattern in the data that is not accounted for by the linear model. Since the Spearman correlation coefficient is not much better than the Pearson coefficient, we do not expect a non-linear fit to be the solution to this problem. We checked this by trying a power-law fit, and indeed found that it was not an improvement. As Table 3 shows, the rms residual of a power-law fit is actually greater than the rms residual of the linear fit; a $\chi^2$ test shows that the residuals for a power law fit are also not normally distributed, while the quantile–quantile plot is even more S-shaped than for the linear fit.

This analysis shows that for the Sun-as-a-star, at the simplest level, a linear model best describes the flux–flux relationship for chromospheric or coronal vs. magnetic flux. Some of the scatter plots of chromospheric flux vs. $|B_{KP}|$, such as that for Mg II *vs.* $|B_{KP}|$ shown in Figure 3, do appear somewhat curved to the eye; however, from the correlation analysis, there is little or no support for fitting a non-linear trend to the data. A power-law fit does not describe the data better than a linear fit, despite the fact that power-law fits have been used in this context by previous studies (*e.g.* Schrijver and Harvey, 1989; Harvey and White, 1999).

**Table 3** Regression coefficients and *rms* residuals for various model fits to the data, $F(\lambda)$.

| $F(\lambda)$ | $R^2$ : linear model fit | | | | $R^2$ : FIR model fit | | | | RMS residual (all data) | | |
|---|---|---|---|---|---|---|---|---|---|---|---|
| | Cycle 21 | Cycle 22 | Cycle 23 | All data | Cycle 21 | Cycle 22 | Cycle 23 | All data | Linear model | Power law model | FIR model |
| S | 0.14 | 0.24 | 0.21 | 0.19 | 0.44 | 0.52 | 0.45 | 0.45 | 0.507 | 0.516 | 0.416 |
| $\Sigma_r$ | - | 0.27 | 0.19 | 0.24 | - | 0.40 | 0.33 | 0.38 | 258 | 274 | 233 |
| $\Sigma_K$ | - | 0.85 | 0.84 | 0.85 | - | 0.89 | 0.89 | 0.90 | 1397 | 1462 | 1158 |
| CaK | - | 0.85 | 0.77 | 0.80 | - | 0.88 | 0.82 | 0.83 | 1.94 $\times 10^{-3}$ | 1.92 $\times 10^{-3}$ | 1.77 $\times 10^{-3}$ |
| Mg II | 0.86 | 0.92 | 0.89 | 0.84 | 0.90 | 0.95 | 0.94 | 0.87 | 2.83 $\times 10^{-3}$ | 2.76 $\times 10^{-3}$ | 2.52 $\times 10^{-3}$ |
| He1083 | 0.77 | 0.89 | 0.85 | 0.74 | 0.83 | 0.93 | 0.90 | 0.78 | 6.37 | 6.59 | 5.88 |
| UV200 | - | 0.85 | - | 0.85 | - | 0.89 | | 0.89 | 0.111 | 0.110 | 0.097 |
| Ly$\alpha$ | 0.78 | 0.89 | 0.82 | 0.79 | 0.84 | 0.94 | 0.89 | 0.84 | 0.381 | 0.405 | 0.333 |
| F10.7 | 0.81 | 0.90 | 0.85 | 0.82 | 0.85 | 0.92 | 0.89 | 0.85 | 23.4 | 24.1 | 21.2 |
| X1 | - | 0.80 | 0.76 | 0.78 | - | 0.85 | 0.84 | 0.84 | 2.7 $\times 10^{-5}$ | 3.4 $\times 10^{-5}$ | 2.3 $\times 10^{-5}$ |
| X2 | - | 0.25 | 0.43 | 0.28 | - | - | - | - | 3.9 $\times 10^{-7}$ | 4.0 $\times 10^{-7}$ | - |

## 3.2 Analysis Using Signal Theory

In Section 3.1, we showed that disk-integrated radiative flux from the solar chromosphere and corona is approximately linearly related to $|B_{KP}|$. But the linear relation cannot account for all of the variance seen in the scatter plots of Figure 2. We now look for an additional temporal relationship between $F(\lambda)$ and $|B_{KP}|$, using the FIR-modeling technique developed by Preminger and Walton (2005). This approach uses signal theory to describe the temporal response of a physical system whose properties are linear and time-independent (*e.g.* Bracewell, 1965).

The "quiet-Sun" state, with no active, magnetic regions present on the solar surface, is assumed to be the equilibrium state of such a system. It is characterized by the quiet-Sun magnetic-flux density [$|B_0|$] and quiet-Sun radiative flux [$F_0(\lambda)$]. The emergence of new magnetic flux constitutes an event that disturbs the equilibrium of the system. In response to the input, $\Delta|B_{KP}|(t) \equiv |B_{KP}|(t) - |B_0|$, the radiative output of the system changes by an amount $\Delta F(\lambda, t)$, such that

$$\Delta F(\lambda, t) \equiv F(\lambda, t) - F_0(\lambda) = \Delta|B_{KP}|(t) * \mathrm{h}_{B,F\lambda}(t) \qquad (1)$$

Here, $*$ denotes convolution, and $\mathrm{h}_{B,F\lambda}(t)$ is called a Finite Impulse Response function (FIR). The FIR describes the temporal response of $F(\lambda)$ to the emergence of magnetic flux.

For all $F(\lambda)$ except X2, we have successfully computed FIRs that describe how each parameter responds to changes in $|B_{KP}|$. We will discuss these FIRs in detail in Section 5. In general, the FIRs have quite a bit of high–frequency noise. This high-frequency noise could be filtered out with a low-pass filter, but not without removing some real high-frequency information, so we have chosen to leave it in. As an example, we present in Figure 3 the FIR for Mg II; the high-frequency noise is clearly visible, especially at large values of $|t|$. The error in the FIR computation depends on the precision of the observations and the validity of the model. We have developed a Monte Carlo method to evaluate the error in each FIR: we take the input and output data sets, shuffle them in a random, but identical, way, and then attempt to compute a FIR using the shuffled data sets and Equation (1). The result is just noise. The standard deviation of the noise so generated is a reasonable estimate of the error in the FIR and is also shown in Figure 3. The FIR signal is significantly larger than the error only for $|t| < 50$ days, so we can consider only this innermost portion of the FIR as representing a real signal.

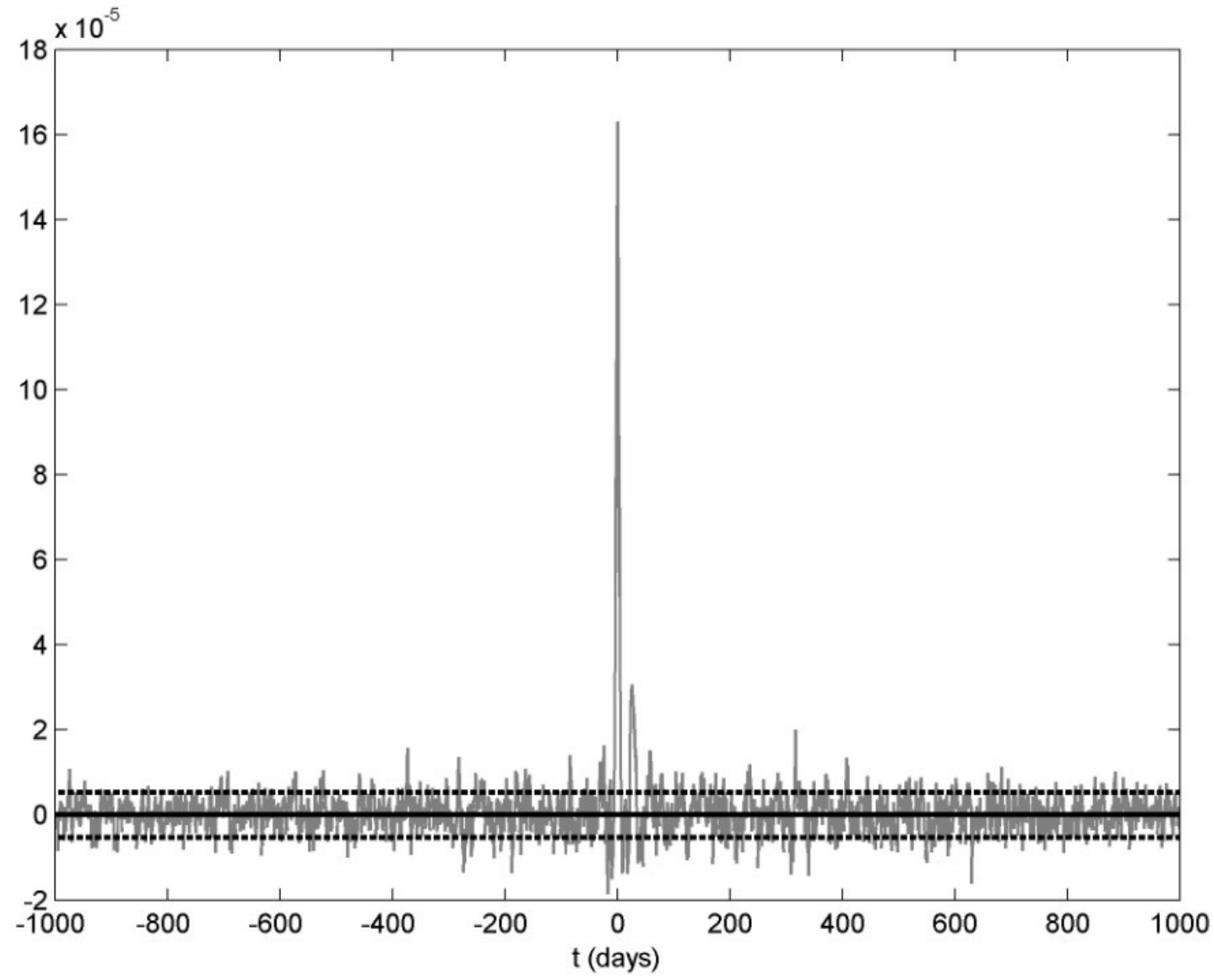

**Figure 3:** FIR function describing the response of Mg II to a brief increase in $|B_{\mathrm{KP}}|$ at $t = 0$. The amplitude of the noise has been estimated using a Monte Carlo method and is indicated by the dashed horizontal lines. Only the portion of the FIR near $t = 0$ has a high signal-to-noise ratio.

In the case of the X2 irradiance, we have tried to use the FIR technique to look for a temporal relation between X2 and $|B_{\mathrm{KP}}|$. While a peak is certainly present at $t = 0$, the FIR signal is so noisy that we do not consider the result to be a well-defined, finite function. This is not entirely unexpected since the reliability of the X2 data is compromised by low sensitivity of the SXT instrument at these wavelengths (L. Acton, private communication, 2009).

In order to reconstruct radiative flux from magnetic-flux density using the FIR model, we use the following procedure: we attenuate each FIR for $|t| > 50$ days, to select the portion of the signal that is significantly greater than the noise, and then use it to reconstruct the change in radiative flux $[\Delta F(\lambda, t)]$ for the given change in magnetic-flux density $[\Delta|B_{\mathrm{KP}}|(t)]$:

$$\Delta F_{\mathrm{rec}}(\lambda, t) = \Delta|B_{\mathrm{KP}}|(t) * \mathrm{h}_{B,F\lambda}(t) \qquad (2)$$

A linear least-squares fit, of the reconstruction $\Delta F_{\mathrm{rec}}(\lambda, t)$ to observations $F(\lambda)$, is needed to recover the value of $F(\lambda)$ in the absence of active regions, $F_0(\lambda)$:

$$F_{\mathrm{obs}}(\lambda, t) = a \cdot \Delta F_{\mathrm{rec}}(\lambda, t) + F_0(\lambda) \qquad (3)$$

For a FIR reconstruction to be considered successful, the fit described by Equation (3) should have a high Pearson coefficient and the coefficient $[a]$ should be close to unity, so that the model characterizes $\Delta F_{\mathrm{rec}}(\lambda, t)$ to within an additive constant. FIR model reconstructions have been carried out for each $F(\lambda)$, and compared with observations via Equation (3); the coefficients of each fit are listed in Table 2.

The coefficients in Table 2 indicate that the FIR method is reasonably successful at modeling the data and the correlation coefficient for the FIR model fit is higher than that for the linear model fit; the improvement is small but it is significant, given the large number of data points involved. In Table 3 we show the regression coefficients $[R^2]$ for the FIR model fit computed cycle by cycle, as well as for all data points. The FIR model fit accounts for more of the variance than the linear model fit in each case. The rms residual is smallest for the FIR model fit. Nevertheless, when a $\chi^2$ goodness-of-fit test is performed on the residuals, we still find that the residuals do not have a normal distribution. A quantile–quantile plot of the residuals is still somewhat S-shaped, albeit less so than was the case for the linear fit. The departure from a normal distribution is particularly pronounced when $F(\lambda)$ is high. In Section 4 we discuss possible causes of this anomaly.

# 4. Reconstructing Radiative Flux from Magnetic-Flux Density

Of the methods tested here, the best reconstruction of radiative flux is obtained using the FIR model. In most cases, when all the data are considered together, the FIR reconstructions predict 85 – 90% of the radiative variability of the chromosphere and corona (See Table 3). In the case of S and $\Sigma_r$, the FIR model fit is only moderately successful. It accounts for the long-term, solar-cycle changes in S and $\Sigma_r$, but is not successful in reproducing the very large short-term dips associated with sunspots moving across the solar disk. Actually, this is not too surprising, because $|B_{KP}|$ does not include extreme dips or spikes (see Figure 1), and convolution is essentially a smoothing procedure; we should not expect the convolution operation of Equation (1) to be able to reproduce sharp variations in the output if none are present in the input. The same is true for X-ray flux: these models can reconstruct the amplitude of solar cycle changes, but not the sharp spikes caused by individual active regions, and/or flares. Figure 4 shows sample observations [$F(\lambda)$] compared with their FIR reconstructions for both a best-case (Mg II) and a worst-case (S) model fit. Also plotted are the residuals for the FIR and (for Mg II) the linear regression model fit. The peak-to-peak variation in the residuals is smaller for the FIR model fit, but the long-term trend in the residuals is similar.

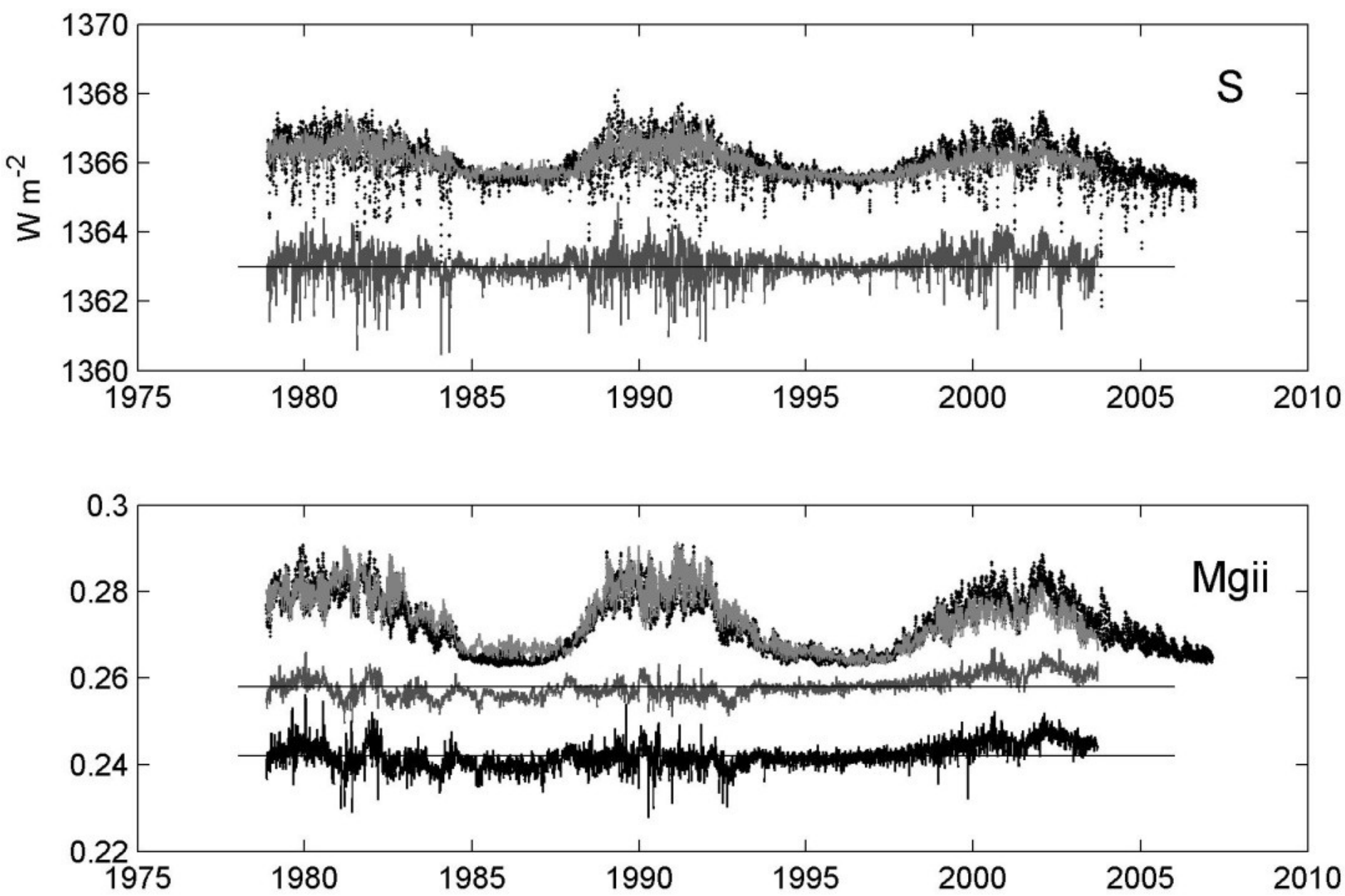

**Figure 4:** Sample observations of radiative flux (dotted curves) and FIR model reconstructions (light gray curves), for both a best-case (Mg II) and a worst-case (S) model fit. Immediately below the observations are the (shifted) residuals for the FIR fits (dark gray curves), and, in the case of Mg II, the (shifted) residual for the linear regression fit (solid black curve). For Mg II, the FIR model accounts for more of the short-term variance, but the long-term trends in the residuals are similar.

Long-term trends and anomalies in the fit residuals may offer some insight into the validity of the model, and may enable us to identify changing patterns of solar activity and/or possible inconsistencies in composite data sets. Figure 5 shows all the residuals of the FIR fits, scaled and shifted so they can be more easily compared to one another. A few features stand out as common to all of the residuals. There is a discontinuity in late 1992, which coincides with the change in the source of $|B_{KP}|$, from the 512-channel magnetograph to the spectromagnetograph at Kitt Peak. The standard deviation of the residuals decreases after this point in time, reflecting the lower noise in the spectromagnetograph measurements. There is a long-term increasing trend in most, but not all, of the residuals during Solar Cycle 23. This trend may explain why the fit residuals are anomalously high when $F(\lambda)$ is high. The fact that this trend is not present in all of the residuals indicates that it is not caused by a trend in $|B_{KP}|$.

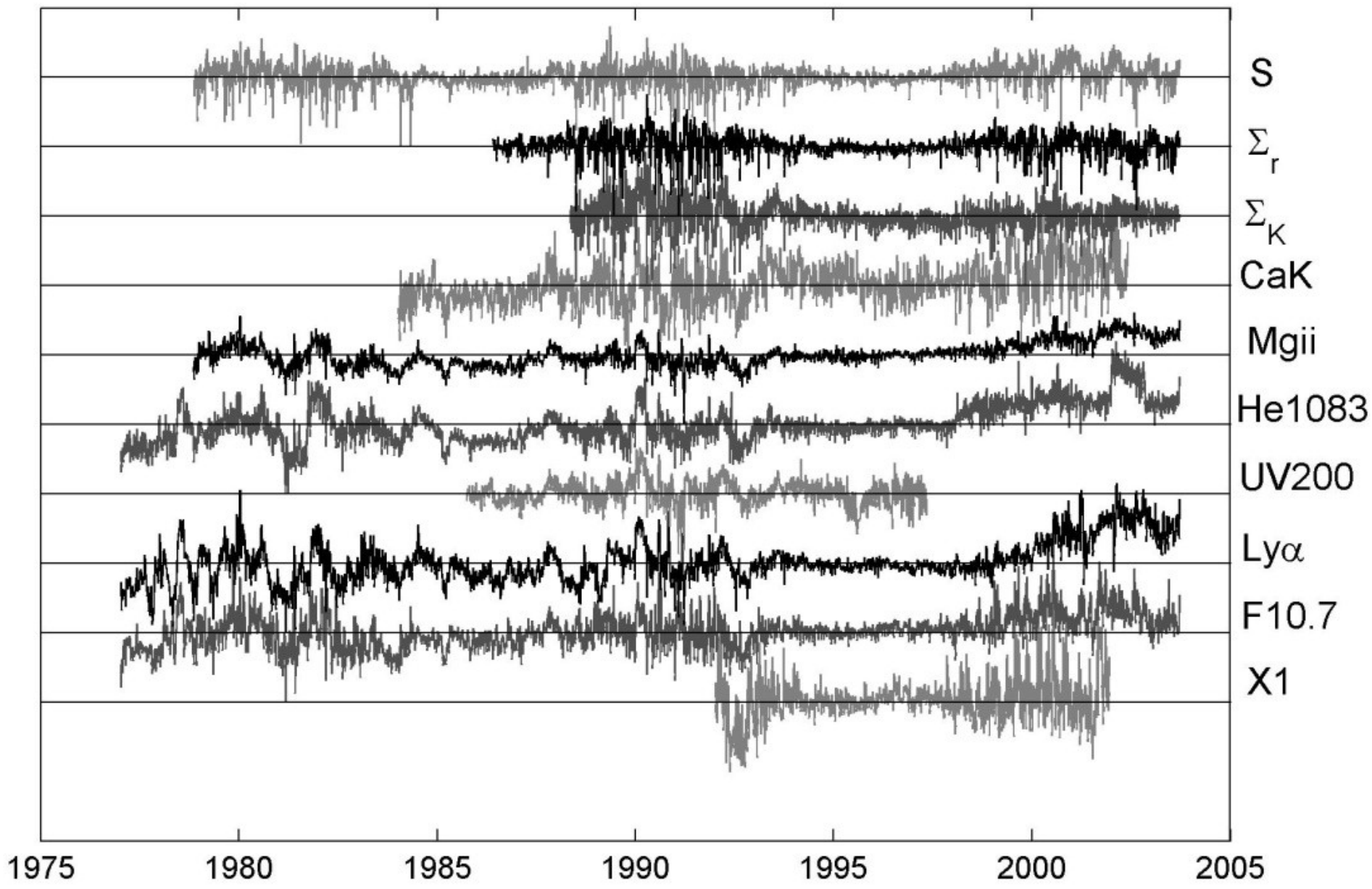

**Figure 5:** Residuals of the FIR model fits, scaled and shifted for comparison purposes.

Many of the measures of solar radiative flux modeled here have previously been modeled using a FIR model based on sunspot area (Preminger and Walton, 2005, 2006; hereafter PW). We cannot compare PW in detail with the new results presented here, because the length of some of the data sequences was greater in PW, and the FIR functions were somewhat smoothed. While the smoothing was an effective way to bring out characteristic patterns in the FIR functions, it also resulted in slightly smoothed reconstructions, and slightly lower regression coefficients for the fits. However, we can compare the results of the two models in general terms. The FIR reconstruction of total solar irradiance [S] is *much* better when the model is based on sunspot area: it can account for 75% of the variance of S, while the FIR model based on $|B_{KP}|$ can only account for 45% of the variance. A comparison of the quantile–quantile plots of the residuals makes it clear that the reason for this difference is the fact that $|B_{KP}|$ seriously underestimates the sunspot component of solar variability. The FIR reconstructions of chromospheric and coronal radiative flux are roughly comparable when the model is based on $|B_{KP}|$ rather than sunspot area. The regression coefficients are similar, while quantile–quantile plots reveal that the distribution of the fit residuals is actually closer to normal when the model is based on $|B_{KP}|$. It is interesting to note that both model fits have similarly high residuals during Cycle 23, for most of the spectral parameters studied. This is unlikely to indicate that the physical process that determines the relationship between radiative and magnetic flux is different from one cycle to another. It may mean that our model omits some source of radiative variability that is more significant during cycle 23 than cycles 21 and 22 (such as the relative amount of small-scale magnetic flux). However, errors in the compilation of the long-term composites for radiative flux cannot be ruled out, because of the difficulties associated with combining observations from two or more instruments, as pointed out by de Toma and White (2006). It is noteworthy that the Cycle 23 residuals for the Ly $\alpha$ reconstruction are very different in this work and PW. We have traced the difference to the use of two different versions of the Lyα composite, published in September and December 2004, respectively.

# 5. Analysis of the FIRs

In Figure 6 we show those portions of the FIRs where the signal-to-noise ratio is high. In this section we discuss, qualitatively, what these FIRs mean. Mathematically, from Equation (1), the FIR function is simply the output of the system when the input is an impulse, *i.e.* the FIR describes the response of radiative flux to a brief increase in magnetic flux. However, we need to make sure that the FIRs are physically reasonable, not just a mathematical construct. From a physical point of view, the temporal behavior of the FIRs starts to make sense when we take into consideration the fact that $|B_{KP}|$ significantly underestimates the flux contributions of sunspots, and small-scale magnetic features (see Section 2).

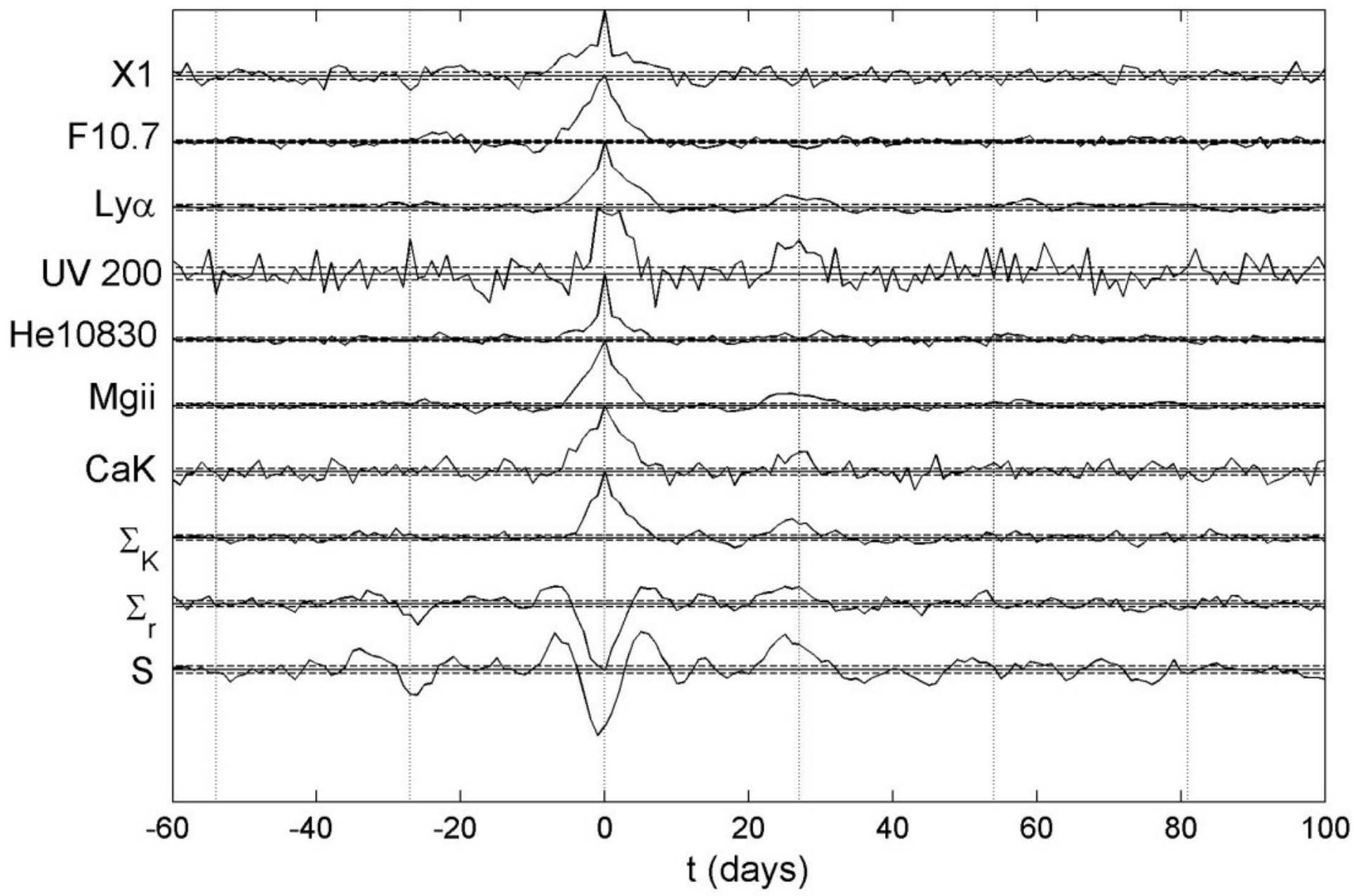

**Figure 6**: FIR functions, scaled to a maximum amplitude of one; zero levels are shifted vertically. Horizontal dashed lines indicate the amplitude of the noise. Vertical dotted lines are drawn every 27 days, for reference.

In Figure 6, vertical lines are drawn every 27 days, about one solar rotation apart. The FIRs all show a strong response to a change in $|B_{KP}|$ at $t = 0$ that is modulated by solar rotation and dies away by $t \approx 50$ days. We expect to see the response at $t = 0$; it describes the radiative response to a magnetic region as it transits the disk.

Notice that the FIRs for various measurements of chromospheric radiative flux also show a small peak at $t = 1$ solar rotation. This is surprising: it means that chromospheric emission remains elevated well after $|B_{KP}|$ has returned to its quiet-Sun state. The physical explanation is that radiative flux from the chromosphere is sensitive to the presence of very small-scale regions that are formed as larger active regions decay. These small regions enhance the chromospheric intensity as the decaying active region rotates into our line of sight for a second time, even though they have become too small to affect $|B_{KP}|$.

The FIRs for X-ray irradiance [X1] from the corona, and F10.7 cm radio flux from the corona and upper chromosphere, do not have a similar peak at $t = 1$ solar rotation. So, coronal radiation, like $|B_{KP}|$, is not sensitive to the smallest decay products of active regions. This is not surprising (Pevtsov and Acton, 2001). What is surprising is the hint of a signal at $t = -1$ solar rotation in the F10.7 FIR, *i.e.* a component that affects the F10.7 flux well before it affects $|B_{KP}|$. This peak is consistent with the negative peaks at $t = -1$ solar rotation for S and $\Sigma_r$, and occurs because $|B_{KP}|$ underestimates sunspot flux, while the F10.7 cm radiation is very sensitive to sunspots. A new sunspot can therefore affect the F10.7 cm flux before it affects $|B_{KP}|$. The FIR for the X-ray radiation, X1, has a sharp spike at $t = 0$. This spike may represent a mathematical way of producing sharp X-ray spikes from the rather smooth $|B_{KP}|$ data.

Also shown in Figure 6 are the FIR functions for S and $\Sigma_r$. These two FIRs show negative as well as positive rotational peaks, according to whether the radiative response is dominated by sunspots (dark) or faculae (bright). The strongest signal is a negative dip at $t = 0$, when the active region is at disk center, flanked by two smaller, positive peaks at $t = \pm 7$ days, when the active region is at the limb. This means that for S and $\Sigma_r$, the radiative contribution of bright faculae outweighs that of dark sunspots when an active region is at the limb. There are two additional signals of significant amplitude, at $t = \pm 1$ solar rotations. The secondary peak at $t = +1$ rotation occurs because there are very small-scale magnetic fields that are decay products of the larger active regions. $|B_{KP}|$ is insensitive to these very small fields, but S and $\Sigma_r$ are affected by them. The dip at $t = -1$ rotation occurs because $|B_{KP}|$ underestimates sunspot flux. S and $\Sigma_r$ are immediately (negatively) affected by a newly emerging sunspot, well before it starts to affect $|B_{KP}|$.

# 6. Summary

We use statistical methods to study the relationship between full-disk solar radiative flux, $[F(\lambda)]$ and average solar photospheric magnetic-flux density $[|B|]$. For a measurement of $|B|$ we use $|B_{KP}|$, the disk-averaged magnetic-flux density observed on Kitt Peak magnetograms, daily from 1977 to 2003, which is a proxy for the total magnetic flux on the visible solar disk. The $F(\lambda)$ are either measurements or proxies of the radiative output from various layers of the solar atmosphere, observed daily for one or more solar cycles.

We use two methods of data analysis: First, we use a statistical correlation and regression analysis to examine the relationship between the instantaneous values of $F(\lambda)$ and $|B_{KP}|$, the so-called flux–flux relations. Total solar irradiance [S] and the variability of red-continuum radiation from the photosphere, $\Sigma_r$, are not monotonic functions of $|B_{KP}|$. However, the radiative flux from the chromosphere and the corona are approximately linear functions of $|B_{KP}|$. The variance in the scatter plots increases with increasing $|B_{KP}|$, which sometimes makes these scatter plots appear curved, but statistical arguments favor a linear fit. This result that the solar flux–flux scatter plots are linear for the chromosphere contradicts earlier results obtained by Schrijver and Harvey (1989), Harvey and White (1999), and Ortiz and Rast (2005); however, none of those studies compared disk-integrated quantities. For coronal radiation, our results are consistent with the results of Pevtsov *et al* (2003) who conclude that coronal X-ray flux is approximately linearly proportional to magnetic flux for other active stars (although their result for the Sun-as-a-star was more ambiguous).

In our second method of data analysis, we apply signal theory to evaluate the temporal nature of the relationship between $F(\lambda)$ and $|B_{KP}|$, using the FIR method introduced by Preminger and Walton (2005). We postulate that $|B_{KP}|$ and $F(\lambda)$ are, respectively, the input and output of a linear physical system and that they are related by convolution with a FIR function (Equation (1)). We find that this model fits the data well and, for each $F(\lambda)$, compute the FIR that describes this relationship explicitly. The FIR functions for radiative emission have a characteristic temporal pattern of peaks depending on whether the radiation arises from the photosphere, chromosphere or corona. The pattern is up to three solar rotations wide.

The empirical models described in this work are used to reconstruct solar spectral irradiance from the observed $|B_{KP}|$. The reconstructions predict 85 – 90% of the radiative variability of the chromosphere and corona, with the FIR model correlating best with observations, according to the goodness-of-fit parameters (Tables 2 and 3). We note that the FIR model reconstructions of the variability of photospheric and total solar irradiance are only moderately successful, because these radiative outputs show strong short-term variability caused by sunspots, a factor that is significantly underestimated by $|B_{KP}|$. The fit residuals are extraordinarily high during Cycle 23 for many measures of radiative flux. This anomaly could be caused by inconsistencies in some of the long-term composite data sets, or by a change in the size distribution of magnetic regions from one solar cycle to the next.

# 7. Conclusions

Our study shows that, for the Sun-as-a-star, the relationship between disk-integrated radiative flux [$F(\lambda)$] and disk-integrated Kitt Peak magnetic flux [$|B_{KP}|$] is essentially linear, but it has a temporal component. Preminger and Walton (2007) showed that when two variables are related in this way, a scatter plot of the variables may not appear linear, even though the underlying physical system is a linear one. The temporal component introduces variance and, sometimes, curvature or a “saturation” effect into the scatter plot.

We have computed FIR functions that describe the temporal nature of the flux–flux relationships explicitly. The temporal pattern in the FIRs lasts for three solar rotations which requires explanation, since previous studies lead us to expect the radiative response of the solar atmosphere to flux emergence to be immediate (*e.g*., Fligge et al, 2000; Krivova et al, 2003; Wenzler et al, 2005; Lockwood, 2005, Figures 48 and 49). In this paper we conclude that the pattern seen in the FIR functions is caused by solar rotation, active region evolution, and the $|B_{KP}|$ underestimation of disk-integrated magnetic flux. Active regions with high magnetic field strength evolve with time, breaking up into small-scale components with low field strength, that affect $F(\lambda)$ and $|B_{KP}|$ differently. In particular, $|B_{KP}|$ significantly underestimates the magnetic flux from sunspots and from small-scale magnetic elements. Radiation from the photosphere and total solar irradiance are affected by all active region components: compact sunspots with extremely intense magnetic fields, large active regions with fields of medium intensity, and diffuse network with weak magnetic fields. Chromospheric emission tends to be most sensitive to large regions and network. Coronal X-ray emission is sensitive to large active regions, while F10.7 cm radiation is sensitive to sunspots and large active regions. This conclusion is consistent with the suggestion by Schrijver (1988) that coronal and chromospheric fluxes are out of phase because chromospheric emission is sensitive to the presence of magnetic network formed from aging, disintegrating active regions, while coronal emission is not. It also explains why a scatter plot of chromospheric *vs.* coronal emission is not linear.

Since this is a Sun-as-a-star study, we expect the results to be applicable to Sun-like stars. The measurement of magnetic fields on cool stars is also subject to instrumental effects. Zeeman-broadened spectral line profiles (used to assess average stellar magnetic-flux density) are primarily sensitive to the magnetic field of bright regions, spots make very little contribution to the line profile because of their low surface brightness (Saar, 1988). Thus, it would be interesting to re-evaluate the flux–flux relations for the chromospheres of Sun-like stars, to see if the non-linearities that have been inferred for these stars could be explained by a temporal component in the underlying relationship, as seems to be the case for the Sun. This is relevant for understanding dynamo action and long-term variability in Sun-like stars (see *e.g.* Nandy, 2004; Nandy and Martens, 2007), specifically where the temporal behavior of the radiative flux is taken as a proxy for magnetic evolution.

We conclude that disk-integrated magnetic flux and radiative flux are related as the input and output of a linear, time-independent physical system. A linear flux–flux relationship is the result that we expect from a disk-integrated study if all active regions are essentially similar, such that two active regions are twice as bright as one. Unfortunately, this result does not give us information regarding the heating mechanism that is at work in any given magnetic active region. A region-by-region study of the flux–flux relationships may or may not be able to provide this important information. Fisher (1998) studied coronal X-ray emission of active regions and found a linear flux–flux relation. We plan to carry out such a study for chromospheric sources of emission. However, since active regions are thought to contain an ensemble of individual flux tube elements (*e.g.* Schrijver *et al*., 1997), high-resolution observations may ultimately be necessary to clarify the flux–flux relationship and lead us toward understanding the mechanism through which magnetic-flux elements heat the solar atmosphere.

**Acknowledgments**

This research has been partially supported by NASA Living with a Star grant NNX07AT19G to Montana State University, by NASA grant NNX08AW53G to the Smithsonian Astrophysical Observatory and by NSF grants ATM-0533511 and ATM-0848518 to the San Fernando Observatory. D.P and G.C. would like to acknowledge the continuing support of CSUN for the program at the San Fernando Observatory. D.N. would like to acknowledge the Department of Science and Technology, Government of India, for support through the Ramanujan Fellowship.

Data from NSO/Kitt Peak are produced cooperatively by NSF/NSO, NASA/GSFC and NOAA/SEC. The composite total solar irradiance data used in this work is version d41_61_0608 from PMOD/WRC, Davos, Switzerland. It includes unpublished data from the VIRGO experiment on the cooperative ESA/NASA Mission SOHO. The daily Ca K emission index is from Sacramento Peak Observatory of the U.S. Air Force Phillips Laboratory. The Mg II core-to-wing index is a composite prepared by the Space Environment Center, Boulder, CO, National Oceanic and Atmospheric Administration (NOAA), US Dept. of Commerce. The 10.7 cm solar radio data are a product of the National Research Council of Canada.